\documentclass[english,twocolumn,english,aps,pre,floats,showpacs]{revtex4-1}
\usepackage[T1]{fontenc}
\usepackage[latin9]{inputenc}
\usepackage{amsmath}
\usepackage{amssymb}
\usepackage{graphicx}
\usepackage{esint}
\usepackage{babel}
\begin{document}

\title{Pulse-noise approach for classical spin systems}

\author{D. A. Garanin}

\affiliation{Physics Department, Lehman College and Graduate School, The City
University of New York, 250 Bedford Park Boulevard West, Bronx, NY
10468-1589, U.S.A.}

\date{\today}
\begin{abstract}
For systems of classical spins interacting with the bath via damping
and thermal noise, the approach is suggested to replace the white
noise by a pulse noise acting at regular time intervals $\Delta t$,
within which the system evolves conservatively. The method is working
well in the typical case of a small dimensionless damping constant
$\lambda$ and allows a considerable speed-up of computations by using
high-order numerical integrators with a large time step $\delta t$
(up to a fraction of the precession period), while keeping $\delta t\ll\Delta t$
to reduce the relative contribution of noise-related operations. In
cases when precession can be discarded, $\delta t$ can be increased
up to a fraction of the relaxation time $\propto1/\lambda$ that leads
to a further speed-up. This makes equilibration speed comparable with
that of Metropolis Monte Carlo. The pulse-noise approach is tested
on single-spin and multi-spin models. 
\end{abstract}

\maketitle

\section{Introduction}

Systems of localized spins on a lattice, considered classically as
vectors $\mathbf{s}$ of unit length, are receiving an increasing
interest because of their application in computer simulations of magnetic
materials. Although this classical model misses the exact form of
the low-temperature magnetization formed by quantum effects (e.g.,
the Bloch law for ferromagnets), it provides a good overall description
of magnetic properties, including non-uniform states. Atomistic models
of classical spins are working at any temperature and overall explain
thermal properties including phase transitions. In this respect they
are superior to \textit{micromagnetics} \textendash{} a macroscopic
approach in which thermal properties have to be taken from experiment. 

The temperature in atomistic classical spin systems can be fixed by
their interaction with the environment modeled by the damping term
introduced by Landau and Lifshitz \cite{lanlif35} plus white-noise-type
random fields $\boldsymbol{\zeta}$ known as Langevin sources and
introduced by Brown \cite{bro63pr}. This equation has the form
\begin{equation}
\mathbf{\dot{s}}=\gamma\left[\mathbf{s}\times\left(\mathbf{H}_{\mathrm{eff}}+\boldsymbol{\zeta}\right)\right]-\gamma\lambda\left[\mathbf{s}\times\left[\mathbf{s}\times\mathbf{H}_{\mathrm{eff}}\right]\right],\label{eq:LLL}
\end{equation}
where $\mathbf{H}_{\mathrm{eff}}$ is the effective magnetic field,
$\gamma$ is gyromagnetic ratio, and $\lambda$ is the dimensionless
damping constant. Such a stochastic model is equivalent to the Fokker-Planck
equation, as was shown by Brown for superparamagnetic particles. The
equilibrium solution of the Fokker-Planck equation should be Boltzmann
distribution, that requires a relation between damping and noise,
\begin{equation}
\left\langle \zeta_{\alpha}(t)\zeta_{\beta}(t')\right\rangle =\frac{2\lambda T}{\gamma\mu_{0}}\delta_{\alpha\beta}\delta(t-t').
\end{equation}
Here $\mu_{0}$ is the magnetic moment of one atom and $k_{B}=1$.
Microscopic theories always suggest $\lambda\ll1$.

It was shown that the vector product in the noise term dictates the
double-vector product form of the Landau-Lifshitz damping term \cite{garishpan90tmf}.
For instance, in the case of noise being anisotropic, the damping
term has to be accordingly modified. Moreover, the main source of
thermal agitation of spins, lattice vibrations, does not produce fluctuating
fields. There are rather fluctuations of the crystal-field anisotropy
tensor. This leads to a more complicated model of noise \cite{garishpan90tmf}
that was not explored, however. The model above, although physically
questionable, is the simplest possible model and it does not lead
to visible inconsistencies.

For many-spin systems, the Fokker-Planck equation becomes a numerically
intractable partial differential equation for the joint probability
density of orientations of all spins. On the other hand, the system
of ordinary differential equations (ODE) of the many-spin stochastic
model can be solved on modern computers in a relatively straighforward
way. Early implementations were done for superparamangetic particles
\cite{libcha93jap,pallaz98prb}. Later the method was applied to magnetic
particles considered as many-spin systems \cite{chubykaloetal02prb,chunowchagar06prb,suhetal08prb,basatxchukac12prb}.
A recent review of the method for magnetic materials provides a link
to a software package developed by authors \cite{evansetal14jpc}.
In particular, the Landau-Lifshitz-Langevin (LLL) method reproduces
the same temperature dependence of the magnetization of the Heisenberg
model, as Metropolis Monte Carlo. 

Currently most of stochastic-dynamics routines for classical spin
systems are using rather primitive Heun method having a quadratic
accuracy in integration steps, the latter being chosen small to avoid
instabilities. For this reason, the computing speeds falls far behind
the speed of the numerical solution of noisless spin models. Generating
noise in these programs is taking longer time than solving equations
of motion.

Fortunately, coupling to the environment $\lambda$ is typically small
in spin systems, so that taking it into account should simplify. The
idea of this work is to replace the white noise by a sequence of pulses
equally spaced in time. In the underdamped case the interval of free
evolution between noise pulses can be made comparable with the spin
precession period or longer, if precession can be discarded, and within
these intervals high-order ODE solvers can be used with a large integration
step.

The main part of this paper is organized as follows. The proposed
pulse-noise method is described in Sec.$\,$\ref{sec:The-method}.
Sec.$\,$\ref{sec:Testing-the-method} is devoted to testing the method
on one-spin problems, including thermally-activated escape. In Sec.$\,$\ref{sec:Many-spin-systems}
the pulse-noise method is tested on many-spin systems and its speed
is compared to that of Monte Carlo.

\section{The method\label{sec:The-method}}

In all existing approaches, noise is considered as invariable within
$n$th integration step $\varDelta t$ and equal to
\begin{equation}
\boldsymbol{\zeta}_{n}=\sqrt{\frac{2\lambda T}{\gamma\mu_{0}\varDelta t}}\mathbf{G}_{n}.\label{eq:zeta_n}
\end{equation}
Here $\mathbf{G}_{n}$ is $n$th realization of a three-component
vector, each component being a normal distribution with a unit dispersion.
Such approximated noise will be called \textit{rectangular} noise.
The coefficient here is fixed by the sum rule
\begin{equation}
\intop_{-\infty}^{\infty}dt\left\langle \zeta_{\alpha}(t)\zeta_{\beta}(t')\right\rangle =\varDelta t\sum_{n}\left\langle \zeta_{n\alpha}\zeta_{n'\beta}\right\rangle =\frac{2\lambda T}{\gamma\mu_{0}}\delta_{\alpha\beta},
\end{equation}
since $\left\langle G_{n\alpha}G_{n'\beta}\right\rangle =\delta_{nn'}\delta_{\alpha\beta}$. 

The usual argument says that invariance of the noise within the integration
step excludes high-order integration methods splitting the step into
several substeps. In this case one has to generate the noise at the
intermediate positions as well, causing a bigger amount of number
crunching. However, since high-order methods require smoothness of
derivatives, they would not work in this case anyway. Nevertheless,
as soon as the rectangular-noise model is already adopted, it is quite
reasonable to solve it with high-order integration methods that are
more accurate and more stable. Below it will be shown that it makes
a considerable positive effect.

The numerical efficiency can be drastically improved by replacing
the rectangular noise by the pulse noise acting only at the boundaries
of intervals $\varDelta t$. The latter can be taken large in the
case of small damping, $\lambda\ll1$. The action of each pulse is
instantaneous rotation of the spin by the angle
\begin{equation}
\boldsymbol{\varphi}_{n}=\sqrt{\varLambda_{N}\varDelta t}\mathbf{G}_{n},\label{eq:phi_n}
\end{equation}
where $\varLambda_{N}\equiv2\gamma\lambda T/\mu_{0}$ is the so-called
Néel attempt frequency \cite{gar97prb}. With $\boldsymbol{\varphi}=\varphi\mathbf{n}$
and $|\mathbf{n}|=1$ the rotation formula reads
\begin{equation}
\mathbf{s}'=\mathbf{s}\cos\varphi+\left(\mathbf{n}\times\mathbf{s}\right)\sin\varphi+\mathbf{n}\left(\mathbf{s\bullet n}\right)\left(1-\cos\varphi\right).\label{eq:s_rotation}
\end{equation}
Such a rotation would occur within the time interval $\varDelta t$
if nothing else than noise acted on the spin. Then, within the intervals
$\varDelta t$, evolution of the noiseless system can be obtained
by an efficient ODE solver making large steps $\delta t$ satisfying
$\delta t<\varDelta t$. The value of $\delta t$ should be chosen
so that noiseless dynamics (mainly precession of spins) be reproduced
correctly. On the other hand, $\varDelta t$ should be a fraction
of the relaxation time due to spin-bath interaction. In the underdamped
case one can choose $\delta t\ll\varDelta t$, drastically reducing
the computer time needed to generate the noise. In this case noisy
dynamics becomes close to the noiseless dynamics, and it is only slightly
modified by random kicks on the spins. 

Separating simultaneous dynamics of the system under the influence
of the noise and everything else into separate motions can be justified
with the help of the Suzuki-Trotter expansion of exponential operators.
The evolution of the system on the time interval $\varDelta t$ can
be represented via the evolution operator $\hat{U}=e^{\hat{A}+\hat{B}}$,
where $\hat{A}$ is due to noiseless dynamics and $\hat{B}$ is due
to noise. Both of these operators depend on $\varDelta t$. In the
underdamped case, if $\varDelta t$ is not too long, $\hat{B}$ becomes
small even if $\hat{A}$ is not. Then one can use the second-order
Suzuki-Trotter formula (see, e.g., Ref.$\,$\cite{hatsuz05Springer})
\begin{equation}
e^{\hat{A}+\hat{B}}\cong e^{\hat{A}/2}e^{\hat{B}}e^{\hat{A}/2}\label{eq:ST-expansion}
\end{equation}
that describes the sequence of (i) noiseless evolution during the
interval $0.5\varDelta t$; (ii) rotation by the noise angle $\boldsymbol{\varphi}$,
Eq.$\,$(\ref{eq:s_rotation}); (iii) repetition of (i). The \textit{a
priori} applicability condition of the pulse-noise approach is 
\begin{equation}
\varphi\sim\sqrt{\varLambda_{N}\varDelta t}\ll1.\label{eq:phi_condition}
\end{equation}
One can see that the weaker is the damping $\lambda$ and the lower
is the temperature, the longer noiseless interval $\varDelta t$ can
be used. It does not make sense to expand Eq.$\,$(\ref{eq:s_rotation})
to the linear order in $\varphi$ since the spin length has to be
conserved. There is another applicability condition, however. Non-thermal
relaxation of the system during the time $\varDelta t$ should be
small,
\begin{equation}
\gamma\lambda H_{\mathrm{eff}}\Delta t\ll1.\label{eq:non-thermal_condition}
\end{equation}
In the opposite case the system will spend most of the time near its
ground state and the averages will correspond to $T=0$. 

All problems described by Eq.$\,$(\ref{eq:LLL}) fall into two categories:
1) Precession term $\gamma\left[\mathbf{s}\times\mathbf{H}_{\mathrm{eff}}\right]$
is important and has to be kept; 2) Precession term can be discarded. 

The first (precessional) case is a regular situation in which using
the pulse-noise model brings a huge computing speed-up in the typical
underdamped case, $\lambda\ll1$. Accurate numerical integration of
the precession imposes the condition $\gamma H_{\mathrm{eff}}\delta t\ll1$,
where $\delta t$ is the integration step used by the ODE solver.
For instance, a good ODE solver provides an acceptable accuracy for
$\gamma H_{\mathrm{eff}}\delta t$ about 0.2 and even 0.25. In this
case, one can use $\Delta t\gg\delta t$ and still satisfy Eqs.$\,$(\ref{eq:phi_condition})
and (\ref{eq:non-thermal_condition}). As the result, the main computer
time is being spent on solving the noiseless dynamics, while the time
spent on computing rare noise kicks becomes negligible. This is true
both for one-spin and many-spin systems.

The second (non-precessional) case is realized if one is interested
in the averages of physical quantities at equilibrium. Since precession
terms do not affect the equilibrium solution of the Fokker-Planck
equation, they can be dropped. If the system has integrals of motion
such as projection the total spin $\mathbf{S}$ on the symmetry axis
$z$ in the case of uniaxial anisotropy, then even the dynamical behavior
of $m_{z}\equiv\left\langle s_{z}\right\rangle $, not only its asymptotic
value, becomes unaffected by the precession. In these cases one can
discard the precession and use the resulting \textit{slow} equation
that can be numerically solved with a much larger integration step
satisfying $\gamma\lambda H_{\mathrm{eff}}\delta t\ll1$. This leads
to an additional speed-up in comparison to the precessional case. 

Within the standard method using continuous noise, dropping the regular
precession term in the equation of motion leads only to a marginal
improvement since there is still the noisy precession term that cannot
be discarded. To the contrast, in the pulse-noise model precession
term can be dropped entierely since precession due to the noise is
accounted for by Eq.$\,$(\ref{eq:s_rotation}).

In the sequel, testing the pulse-noise approach will be done for different
models. For the sake of transparency, instead of introducing different
reduced quantities, some parameters and constants, such as $\gamma$,
$\mu_{0}$, etc., will be set to 1. This should be taken into account
in reading plots. 

Details of the numerical implementation using Wolfram Mathematica
are given in the Appendix.

\section{Testing the method on one-spin problems\label{sec:Testing-the-method}}

\subsection{Spin in a magnetic field}

\begin{figure}
\begin{centering}
\includegraphics[width=8cm]{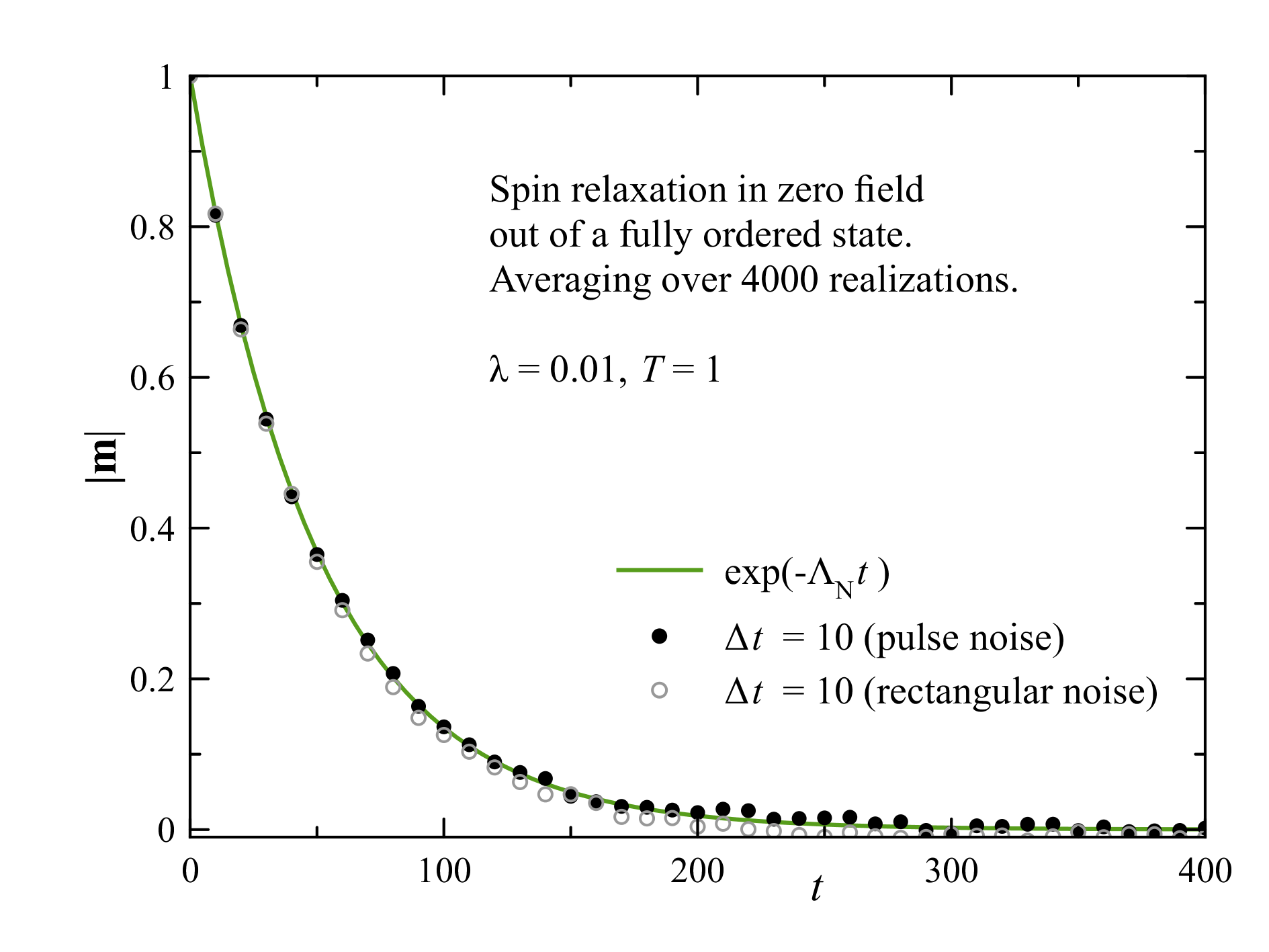}
\par\end{centering}
\caption{Exponential decay of the magnetization out of a completely ordered
state under the action of the noise only (the high-temperature limit).
The pulse-noise and rectangular-noise models yield the same results.}

\label{Fig_relax_szAvr_vs_t_Hz=00003D0_T=00003D1_tRel=00003D100_tFree=00003D10}
\end{figure}

The first test to make is the test of the discretization of the noise
into intervals $\varDelta t$ for the trivial system in which only
the noise is present, that physically corresponds to high temperatures.
In this case the Fokker-Planck equation for one spin readily yields
the evolution of the magnetization $\mathbf{m}=\left\langle \mathbf{s}\right\rangle $
in the form
\begin{equation}
\mathbf{m}=\mathbf{m}_{0}e^{-\varLambda_{N}t}.
\end{equation}
Fig.$\,$\ref{Fig_relax_szAvr_vs_t_Hz=00003D0_T=00003D1_tRel=00003D100_tFree=00003D10}
shows this dependence together with the results of numerical solutions
of the pulse-noise and rectangular-noise models with $\delta t=0.25$
and discretization time $\Delta t=10$ for $\lambda=0.01$, $T=1$.
Here the RMS value of the rotation angle $\varphi_{\mathrm{RMS}}=\sqrt{\varLambda_{N}\varDelta t}\simeq0.45$
is not small, still the pulse-noise model well reproduces the analytical
result. The model with rectangular noise yields the same result that
is not surprising. Whereas in the pulse-noise model rotations are
instantaneous in the middle of the $\Delta t$ interval, in the rectangular-noise
model they are performed gradually by the ODE solver to the same effect.

\begin{figure}
\begin{centering}
\includegraphics[width=8cm]{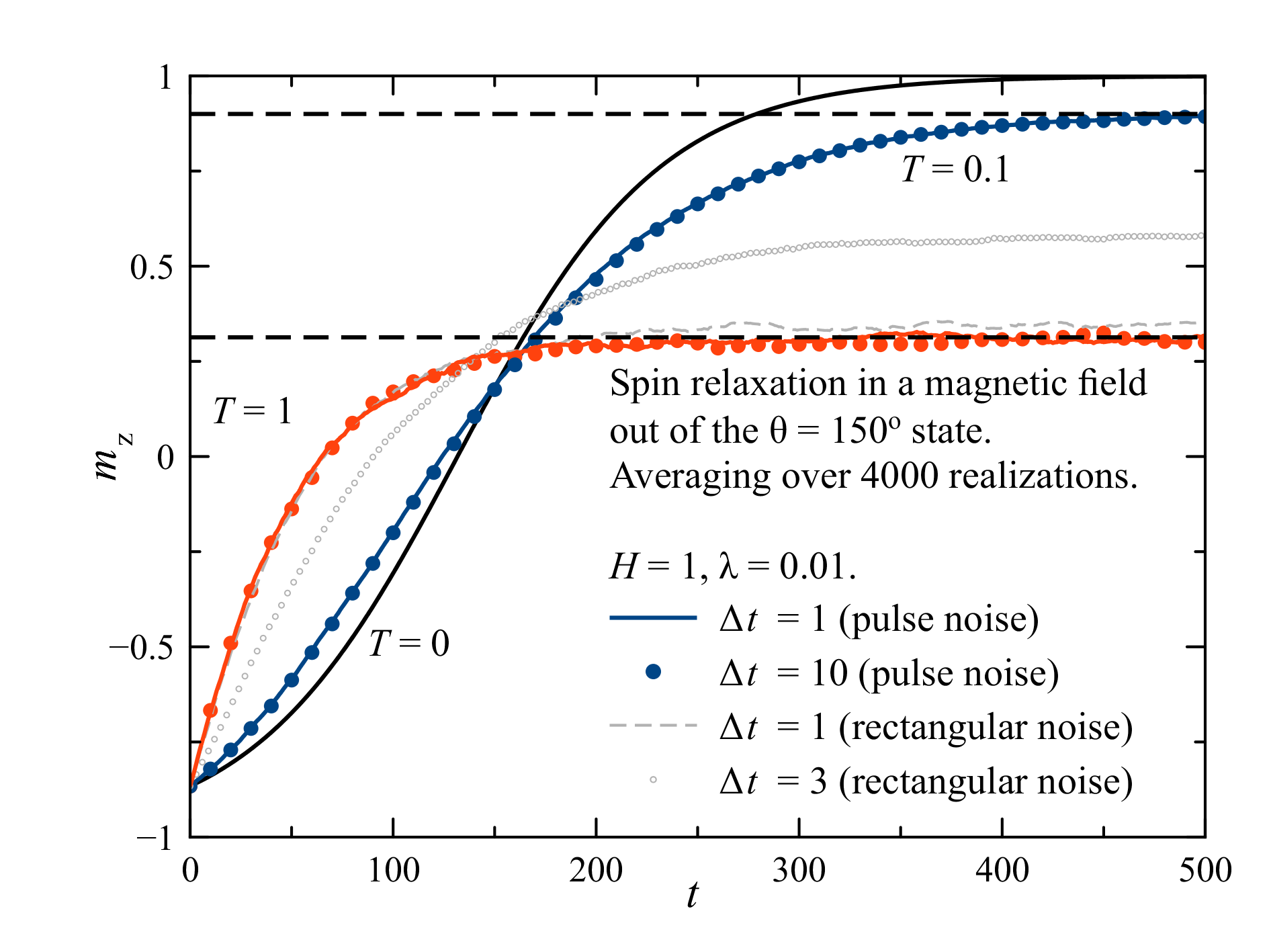}
\par\end{centering}
\caption{Relaxation of the spin in a magnetic field out of the completely ordered
state directed at $\theta=150\protect\textdegree$ to the field, at
different temperatures and $\varDelta t$. Pulse-noise model works
perfectly for $\Delta t=1$ (red, blue, and black lines) and holds
up to $\Delta t=10$ (red and blue points), while the rectangular-noise
model (gray) fails already at $\Delta t=3$.}

\label{Fig_relax_szAvr_vs_t_theta=00003D150_Hz=00003D1_tRel=00003D100}
\end{figure}

For non-trivial spin Hamiltonians, starting with the spin in a magnetic
field, the difference between the two noise models becomes tremendous.
This can be seen in Fig.$\,$\ref{Fig_relax_szAvr_vs_t_theta=00003D150_Hz=00003D1_tRel=00003D100},
where initially the spin is directed at $\theta=150^{\circ}$ to the
field. Whereas the pulse-noise model yields visually the same results
for $m_{z}(t)$ for $\Delta t=1$ and 10 (both with $\lambda=0.01$
and $H=1$) and asymptotically approaches the correct equilibrium
value, the rectangular-noise model for $T=1$ is working only for
$\Delta t=1$, although there a visible overestimation of $m_{z}(\infty)$.
Already for $\Delta t=3$ this model breaks down completely, mimicking
a significantly lower temperature. This can be interpreted as the
rectangular noise being correlated and thus gentle, only slightly
modifying the field instead of really kicking the spin. Here in all
cases RK5 ODE solver with the integration time step $\delta t=0.25$
was used. For the sake of comparison, precession term was kept in
the equation of motion, since for the rectangular-noise model it cannot
be discarded. 

The spin-in-a-field model is convenient for making a comparison with
the standard stochastic-dynamics approach using the Heun ODE solver.
It was found that for $T=1$ and other parameters as indicated above,
the Heun solver is stable for $\delta t\leq0.04$, where it yields
visibly same results as the pulse-noise method in Fig.$\,$\ref{Fig_relax_szAvr_vs_t_theta=00003D150_Hz=00003D1_tRel=00003D100}.
Above this value the Heun method crashes even if the spin length is
constantly corrected. Ref.$\,$\cite{evansetal14jpc} uses natural
units with $J=3\times10^{-21}$ J/link for magnetic particles. The
time step was $\delta t_{\mathrm{natural}}=10^{-15}$s but it had
to be decreased to $10^{-16}$s near the Curie point. In dimensionless
units used here, $10^{-15}$s corresponds to $\delta t=\left(J/\hbar\right)\delta t_{\mathrm{natural}}\simeq0.03$.
Other authors also report using rather small time steps with the Heun
method, that makes it slow. Within the Heun method in the present
implementation, most of the computer time is being spent on generation
of random numbers, and the resulting computing speed is 10 times lower
than that of the pulse-noise method using RK5 solver with $\delta t=0.25$
and $\Delta t=10$. 

Even within the rectangular-noise model, using the more stable RK5
instead of the Heun method allows to use $\delta t=0.25$ to reach
a speed-up by a factor of 4. This confirms the statement about usefulness
of high-order integration methods made at the beginning of Sec.$\,$\ref{sec:The-method}.

\begin{figure}
\begin{centering}
\includegraphics[width=8cm]{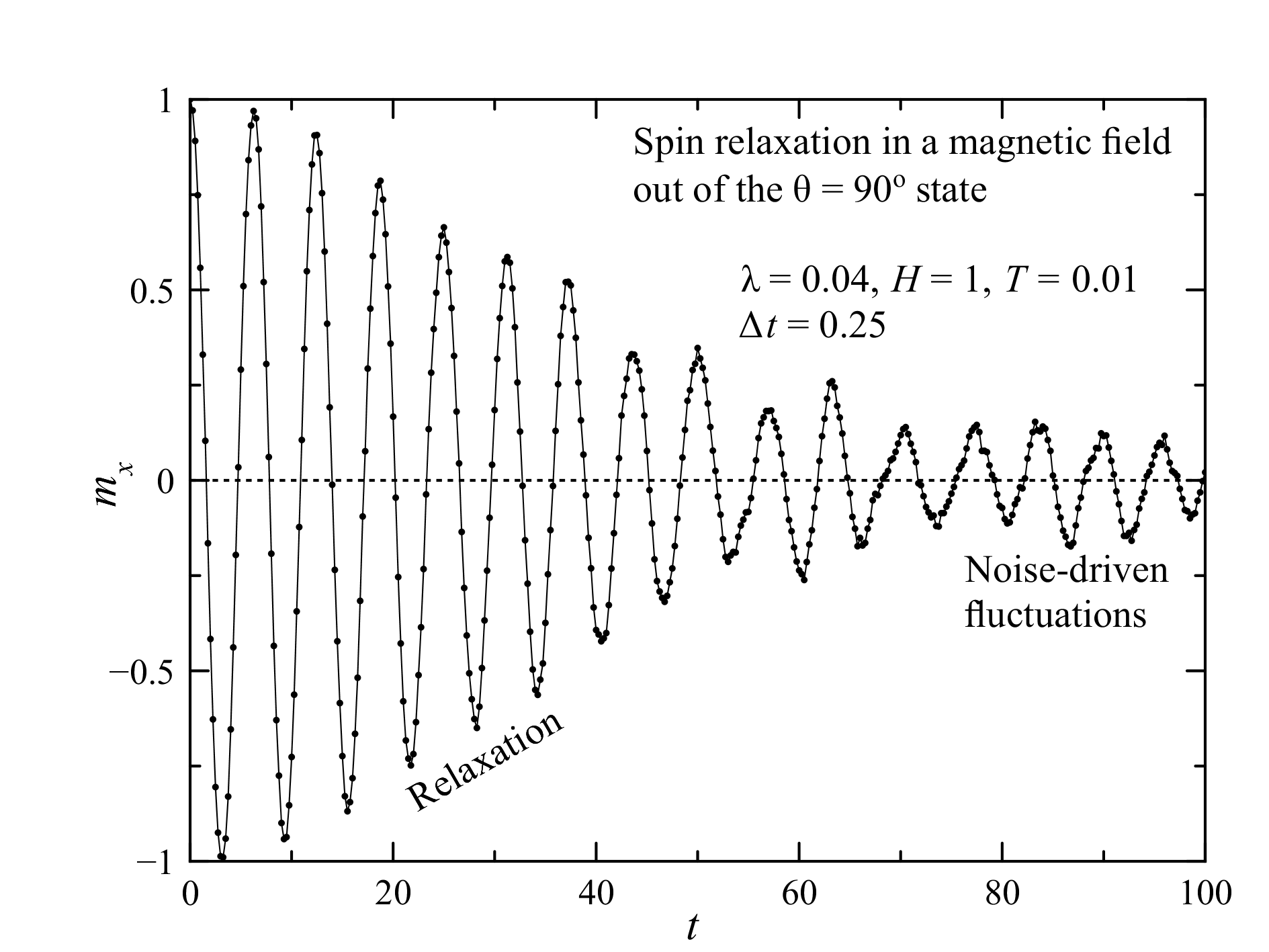}
\par\end{centering}
\caption{One realization of the spin precession with relaxation in the presence
of a pulse noise. The initial direction of the spin is perpendicular
to the field. For the temperature as low as $T=0.01,$ there is a
regime of nearly-deterministic relaxation terminating in the regime
of noise-driven fluctuations.}
\label{Fig_precession_sx_vs_t_theta=00003D90_Hz=00003D1_T=00003D0d01_tRel=00003D25_tFree=00003D0d25}
\end{figure}

Fig.$\,$\ref{Fig_precession_sx_vs_t_theta=00003D90_Hz=00003D1_T=00003D0d01_tRel=00003D25_tFree=00003D0d25}
shows one realization of the spin precession with relaxation in a
magnetic field in the presence of a pulse noise for $H=1$, $T=0.01$,
$\lambda=0.04$. Here the computation was done with $\delta t=0.25$
and $\varDelta t=\delta t$. One can see precession with relaxation
terminating in a noisy behavior. 

Note that noiseless precession and relaxation of a spin in a magnetic
field can be described analytically, so that there is no need of numerical
integration. Transformation of $s_{z}$ and the transverse spin component
$s_{\bot}=\sqrt{s_{x}^{2}+s_{y}^{2}}$ during the time interval $\Delta t$
is described by \cite{chugar02prl}
\begin{eqnarray}
s_{z}' & = & \frac{\sinh\left(\tau\right)+s_{z}\cosh\left(\tau\right)}{\cosh\left(\tau\right)+s_{z}\sinh\left(\tau\right)}\nonumber \\
s_{\bot}' & = & \sqrt{1-s_{z}'^{2}},\qquad\tau\equiv\gamma\lambda H\Delta t.\label{eq:s_evolution_anal}
\end{eqnarray}
Here one can trivially add precession to find the values of $s_{x}$
and $s_{y}$. Thus evolution of the spin in a field within the pulse-noise
model is a map combined of discrete transformations of two kinds.
Same is true for the rectangular-noise model, although working out
analytics is more cumbersome because of changing the direction of
the total field. Although the transformation above is exact, $\tau$
has to be small because of Eq.$\,$(\ref{eq:non-thermal_condition}).

Let us now investigate the \textit{a posteriori} accuracy of the pulse-noise
approximation by looking at the equilibrium value of $m_{z}$ obtained
by extensive averaging for the spin-in-a-field model using Eqs.$\,$(\ref{eq:s_rotation})
and (\ref{eq:s_evolution_anal}). After an initial thermalization
period, spin evolution was monitored within the time interval $t_{\max}=10^{6}$,
and $m_{z}=\left\langle s_{z}\right\rangle $ was computed by averaging
the values at the end of each time interval $\Delta t$. Such computations
were run in parallel cycles, using three different computers having
4, 8, and 16 cores. The final computed average corresponds to the
total averaging time $t_{\mathrm{avr}}=t_{\max}\times N_{\mathrm{cores}}\times N_{\mathrm{cycles}}$. 

\begin{figure}
\begin{centering}
\includegraphics[width=8cm]{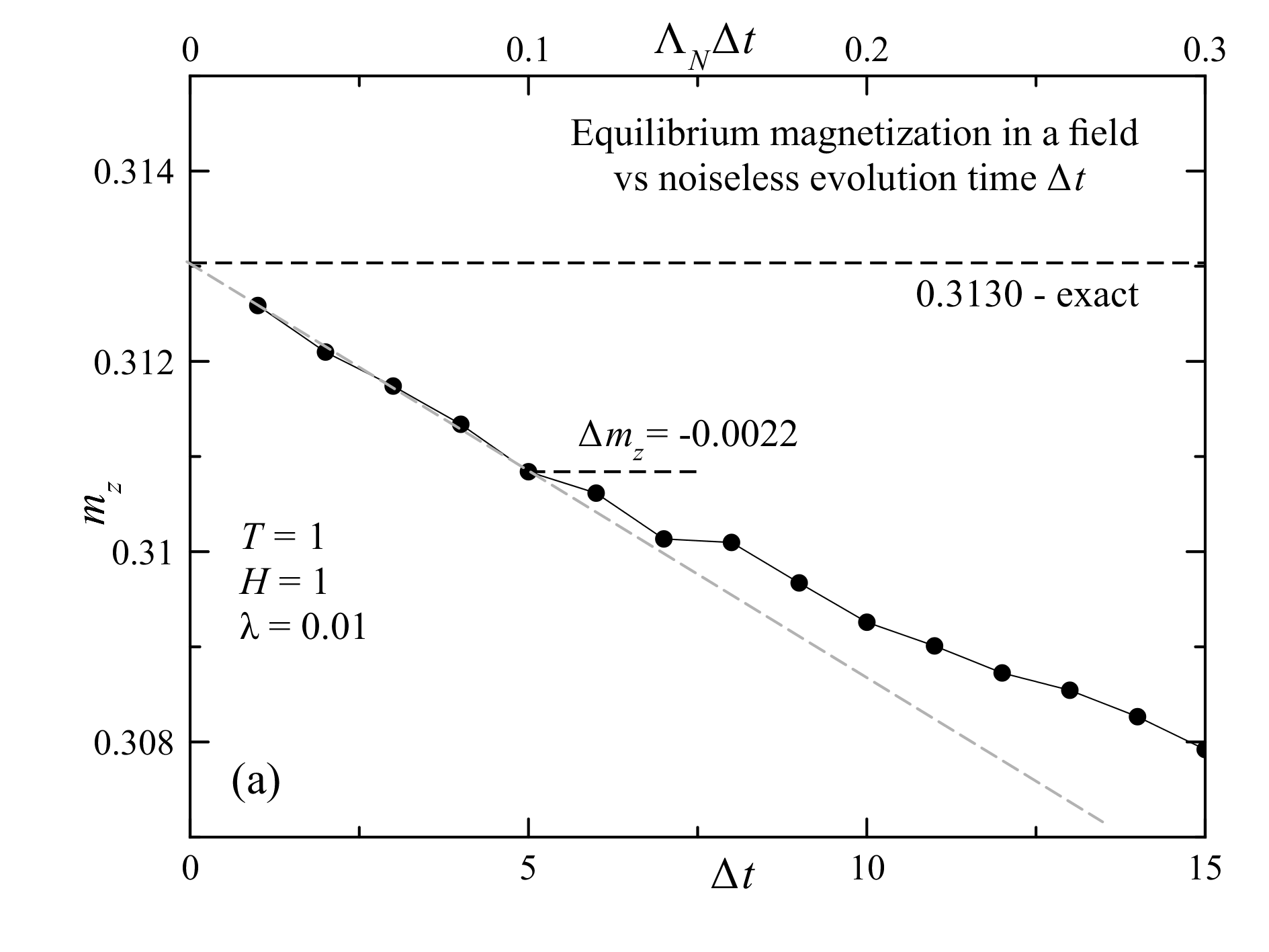}
\par\end{centering}
\begin{centering}
\includegraphics[width=8cm]{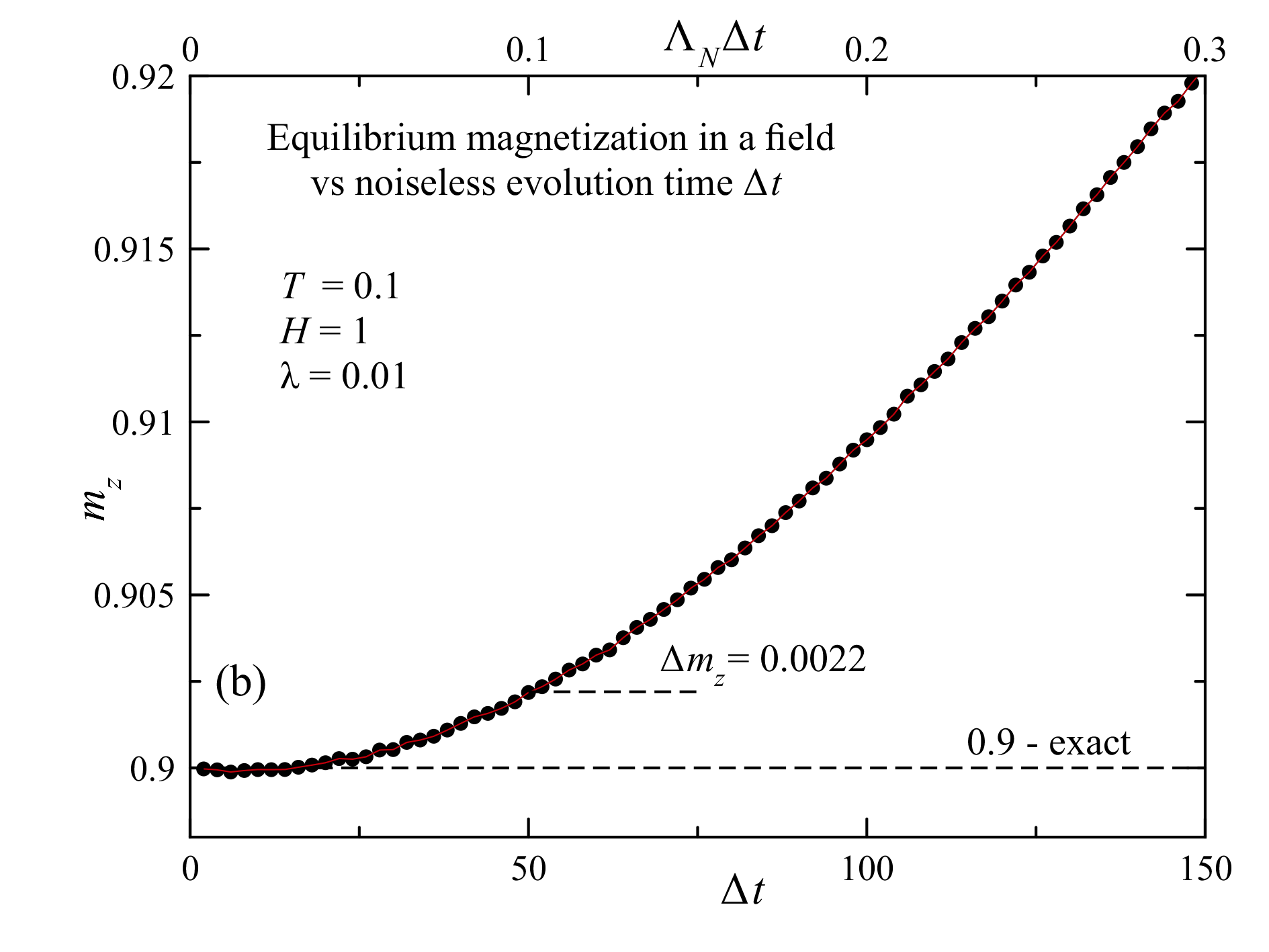}
\par\end{centering}
\caption{Equilibrium magnetization $m_{z}$ for the spin-in-a-field model vs
noiseless evolution time $\Delta t$ (bottom $x$ axes) and parameter
$\varLambda_{N}\Delta t$ (top $x$ axes) for $H=1$ and $\lambda=0.01$.
The curves begin from the exact values of $m_{z}$ on the left, and
the error of the pulse-noise approximation increases with $\varDelta t$.
(a) $T=1$; (b) $T=0.1.$ }

\label{Fig_szAvr_vs_tFree}
\end{figure}

Fig.$\,$\ref{Fig_szAvr_vs_tFree} shows the dependence of $m_{z}$
computed with $t_{\mathrm{avr}}\simeq2\times10^{9}$, as explained
above, on $\Delta t$ for $H=1$ and $\lambda=0.01$. At the elevated
temperature $T=1$ (Fig.$\,$\ref{Fig_szAvr_vs_tFree}a), the deviation
from the exact result goes down almost linearly with some upward curwature.
This upward curvature is dominating at the low temperature $T=0.1$
in Fig.$\,$\ref{Fig_szAvr_vs_tFree}b, so that the deviation from
the exact result is positive. This can be explained by the effect
commented upon below Eq.$\,$(\ref{eq:non-thermal_condition}), since
here $\gamma\lambda H\Delta t=1$ at $\Delta t=100$. However, the
effect is smaller than expected, thus the applicability condition
in Eq.$\,$(\ref{eq:non-thermal_condition}) is somewhet less stringent
than it seems. The scales of $\Delta t$ in both computations were
chosen so that the range of $\varLambda_{N}\Delta t$ is the same,
as shown in top $x$ axes. In both cases $\varLambda_{N}\Delta t=0.1$
provides an accuracy good enough, as shown in the figures, and it
satisfies the thermal applicability condition, Eq.$\,(\ref{eq:phi_condition}).$
Note that low temperatures are more favorable for the pulse-noise
model: $\varLambda_{N}\Delta t=0.1$ corresponds to $\Delta t=5$
for $T=1$ and $\Delta t=50$ for $T=0.1$. In the plots, $\varDelta m_{z}$
is the deviation from the exact value of $m_{z}$, that for $\varLambda_{N}\Delta t=0.1$
has different signs for $T=1$ and $T=0.1$.

\subsection{Thermally activated escape rate of a uniaxial spin in a transverse
field\label{subsec:Thermal-activation-escape}}

\begin{figure}
\begin{centering}
\includegraphics[width=8cm]{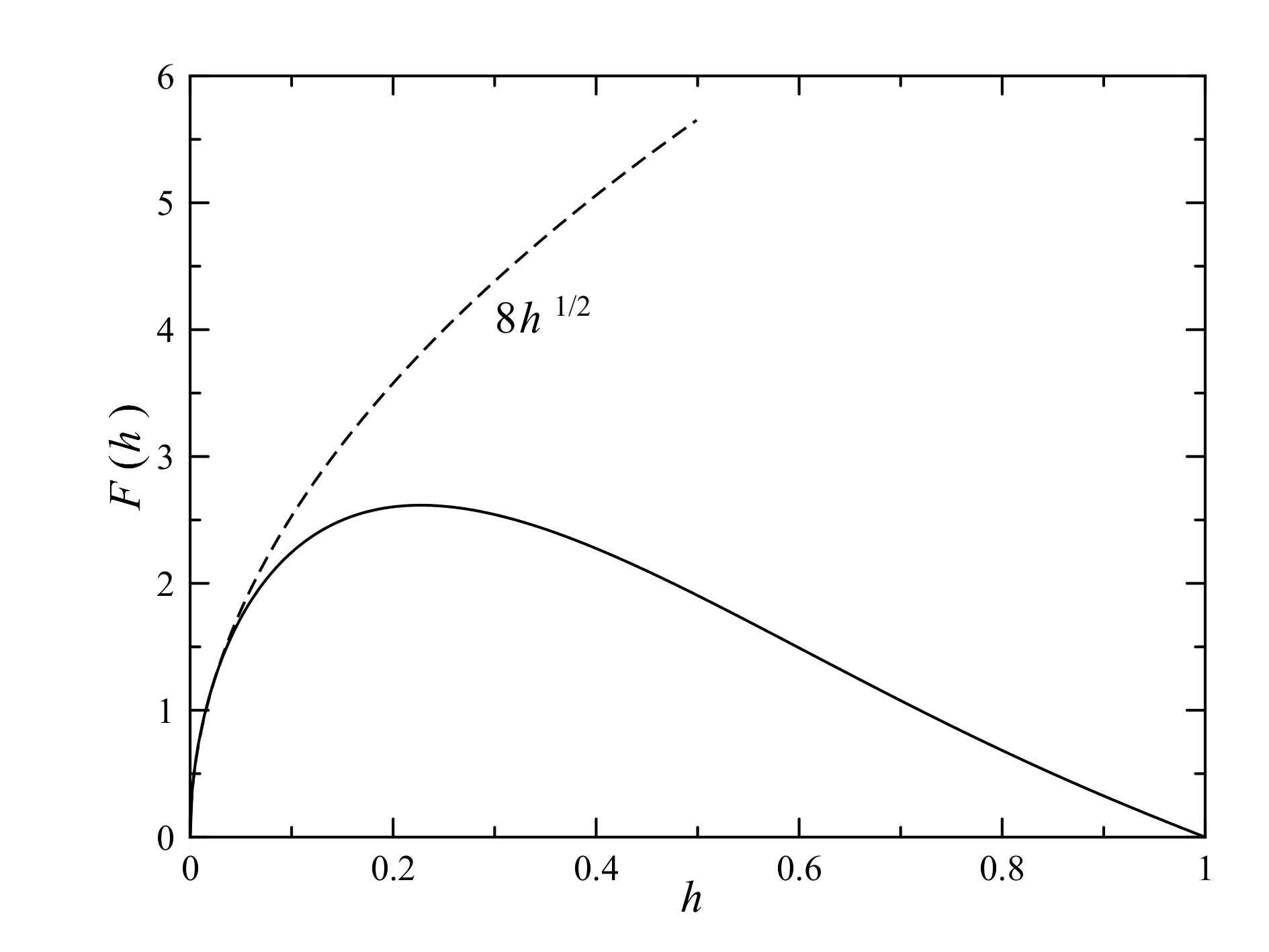}
\par\end{centering}
\caption{Function $F(h)$ in the low-damping escape rate of a uniaxial spin
in a transverse field, Eq.$\,$(\ref{eq:ALD}).}

\label{Fig_FLD_vs_h}
\end{figure}

Uniaxial spin in a transverse field is an example of a system, for
which precession is relevant in the dynamics. The energy 
\begin{equation}
\mathcal{H}=-Ds_{z}^{2}-Hs_{x}
\end{equation}
for $h\equiv H/(2D)<1$ possesses two degenerate minima at the angle
$\theta=\arcsin h$ to $z$ axis. The saddle point is $\mathbf{s}=(1,0,0)$
and the energy barrier between the minima is given by $\Delta U=D(1-h)^{2}$.
In the case of a well-developed saddle, the thermally activated escape
rate over the barrier $\Gamma$ reads \cite{garkencrocof99pre} 
\begin{equation}
\Gamma=\frac{\omega_{0}}{2\pi}Ae^{-\sigma(1-h)^{2}},\label{eq:GammaEsc}
\end{equation}
where $\sigma\equiv D/T$ and $\omega_{0}=2\gamma D\sqrt{1-h^{2}}$
is the frequency of the ferromagnetic resonance near the bottom of
the well, so that $\omega_{0}/(2\pi)$ can be interpereted as the
\textit{attempt frequency}. The factor $A$ has different forms in
the high-damping (HD), intermediate damping (ID) and low damping (LD)
regimes, similarly to the problem of a particle in a potential well
considered by Kramers \cite{kra40}. Crossovers between these regimes
and those to the uniaxial case have been studied in Ref.$\,$\cite{garkencrocof99pre}.
In the HD regime, $\lambda\gtrsim1$, one has $A\propto\lambda$ (or
$A\propto1/\lambda$ if the Gilbert equation is used). Since HD regime
is untypical for spin systems, it will not be considered here. In
the ID regime $\lambda\lesssim1$ that corresponds to the \textit{transition-state
theory}, one has $A=1$. Finally, in the LD case the energy dissipated
over the separatrix trajectory around one well becomes smaller than
thermal energy, $\delta E\sim\lambda D\lesssim T$, and the \textit{energy-diffusion
regime} sets in. In this case one has $A=\delta E/(2T)$ that can
be written in the form
\begin{equation}
A\equiv A^{(LD)}=\lambda\sigma F(h),\label{eq:ALD}
\end{equation}
where
\begin{equation}
F(h)=\frac{1}{2}\oint_{f=f_{c}}\left[(1-x^{2})\frac{\partial f}{\partial x}d\phi-\frac{1}{1-x^{2}}\frac{\partial f}{\partial\phi}dx\right],
\end{equation}
$x\equiv\cos\theta$, and $f$ is the dimensionless energy in the
spherical coordinates,
\begin{equation}
f(x,\phi)=-\mathcal{H}/D=x^{2}+2h\sqrt{1-x^{2}}\cos\phi,
\end{equation}
 $f_{c}=2h$. The maximal value of $x$ on the separatrix is given
by $x_{c}=2\sqrt{h(1-h)}$. It is convenient to calculate $F(h)$
as the integral over $x$ over the half of the separatrix between
0 and $x_{c}$, with $\phi=0$ at both points. After some algebra
one obtains 
\begin{equation}
F(h)=\intop_{0}^{1}\frac{x_{c}du}{1-x_{c}^{2}u^{2}}\left[\frac{u\left[2(1-h)-x_{c}^{2}u^{2}\right]^{2}}{\sqrt{1-u^{2}}}+x_{c}\sqrt{1-u^{2}}\right]
\end{equation}
that can be computed numerically, see Fig.$\,$\ref{Fig_FLD_vs_h}.
For $h\ll1$ this simplifies to $F(h)=8\sqrt{h}$ \cite{kligun90,garkencrocof99pre}.

Non-trivial crossover between the ID and LD regimes is given by the
Melnikov's formula \cite{mel85,melmes86,cofgarcar01acp}. However,
for the current purposes (plotting the escape rate in a log scale)
it is sufficient to use the interpolation
\begin{equation}
A=A^{(ILD)}=\frac{A^{(LD)}}{1+A^{(LD)}}.\label{eq:AILD_interpolation}
\end{equation}

\begin{figure}
\begin{centering}
\includegraphics[width=8cm]{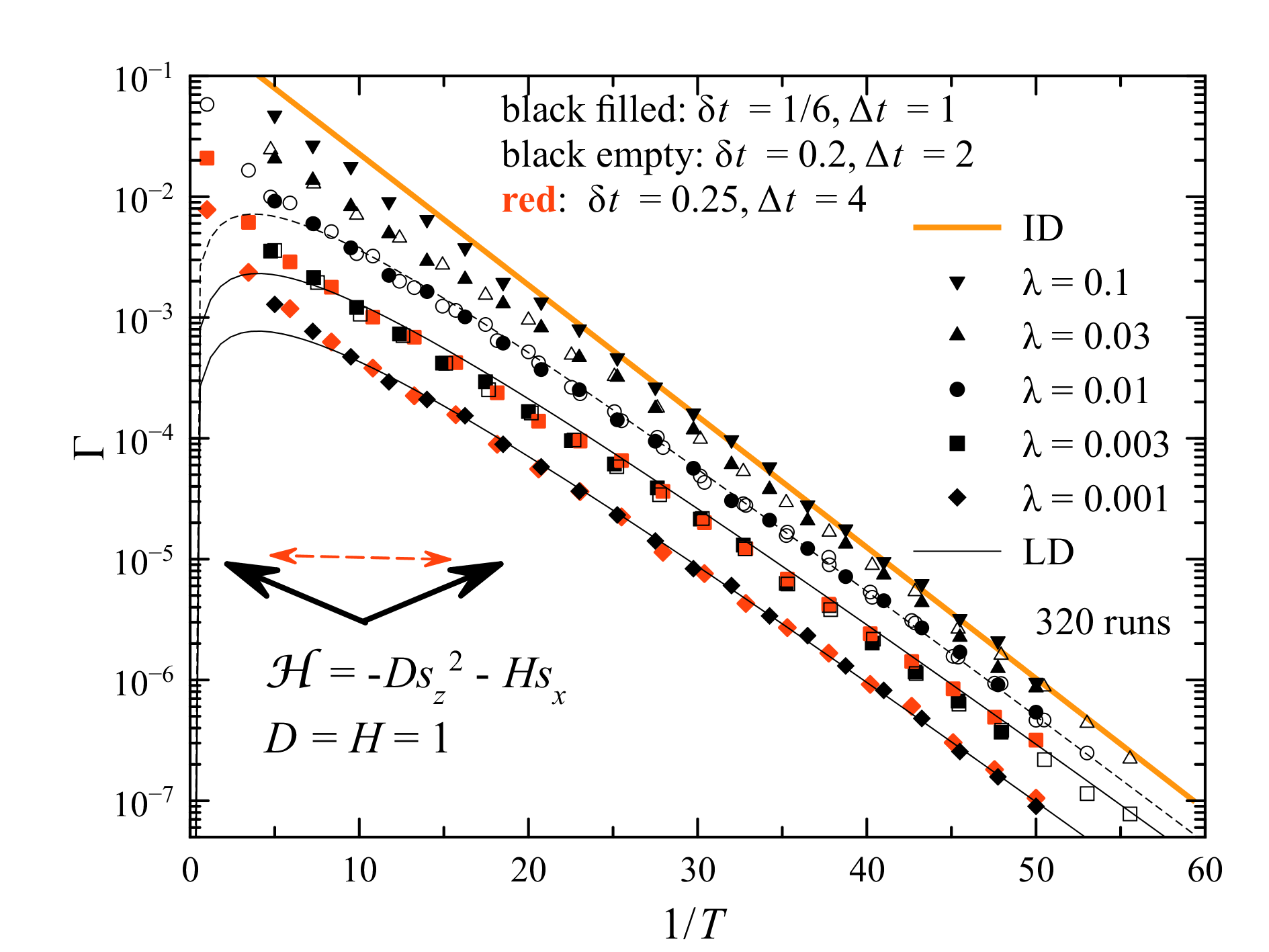}
\par\end{centering}
\caption{Temperature dependence of the thermal activation escape rate of a
uniaxial spin in a transverse field for different values of damping
$\lambda$, obtained with different values of the parameters in the
pulse-noise approximation. The ID line corresponds to $A=1$ in Eq.$\,$(\ref{eq:GammaEsc}).
The LD lines use Eq.$\,$(\ref{eq:ALD}) with $F(h)=1.90$ for $h=1/2$.
The dashed line uses the LD-ID interpolation, Eq.$\,$(\ref{eq:AILD_interpolation}).}

\label{Fig_GammaEsc_vs_Tinv_D=00003D1_Hx=00003D1_tFree=00003D4_hStep=00003D0d25}
\end{figure}

This problem has been investigated by different methods (see, e.g.,
Ref.$\:$\cite{kalmykovetal10prb} and references therein). The results
of numerical calculation of the escape rate using Eq.$\,$(\ref{eq:LLL})
with the pulse-noise approximation with different parameters are shown
in Fig.$\,$\ref{Fig_GammaEsc_vs_Tinv_D=00003D1_Hx=00003D1_tFree=00003D4_hStep=00003D0d25}
together with analytical results with which they fully agree. In the
computations, spins were initially put at the bottom of the potential
well $\phi=0$, $\sin\theta=h$ with $h=1/2$ and then the evolution
routine was run using Butcher's RK5 ODE solver with time step $\delta t$
and random pulse rotations with noise-free evolution time $\Delta t$.
By construction {[}see Eq.$\,$(\ref{eq:ST-expansion}){]} the ratio
$\Delta t/\delta t$ has to be an even number. After crossing the
line $s_{z}=0$ the computation was stopped and the first-passage
time was recorded. For each temperature 320 runs were done in parallel,
the mean first-passage time (MFPT) was computed and escape rate $\Gamma$
was found as its inverse. A similar procedure is being used in the
experiment \cite{cofetal98prl}.

The results in Fig.$\,$\ref{Fig_GammaEsc_vs_Tinv_D=00003D1_Hx=00003D1_tFree=00003D4_hStep=00003D0d25}
show that for $\lambda=0.1$ the ID regime, Eq.$\,$(\ref{eq:GammaEsc})
with $A=1,$ is realized for most temperatures. On the other hand,
the results for $\lambda=0.003$ and 0.001 are well described by the
LD formula, Eq.$\,(\ref{eq:ALD})$. The cases $\lambda=0.03$ and
0.01 are ILD crossover cases. In particular, the $\lambda=0.01$ results
are well described by the interpolation formula, Eq.$\,$(\ref{eq:AILD_interpolation}).

Concerning the accuracy of computations, the set $\delta t=1/6$ and
$\Delta t=1$ was used as the reference one as it provides accurate
resuls for all dampings and temperatures studied here. Already for
this set, the ratio $\Delta t/\delta t=6$ ensured that the computer
time spent on generating random numbers and rotations of the spin
is negligibly small in comparison to the time spent on solving the
noiseless equation of motion. For higher values of damping, $\lambda=0.01$
and 0.03, the computation could be sped up by choosing $\delta t=0.2$
and $\Delta t=2$ with essentially the same results. However, for
$\Delta t=4$ obtained values of $\Gamma$ were visibly too high.
This can be explained by strong kicks allowing spins to cross the
barrier at once from a position slightly below it, that results in
effective reducing the barrier. For lower damping, such as $\lambda=0.001$
and 0.003, the set $\delta t=0.25$ and $\Delta t=4$ could be used
without significant loss of accuracy, that allowed an even greater
speed-up. Increasing integration step above $\delta t=0.25$ leads
to a sharp decrease of accuracy and even to an instability. Thus the
integration time step larger than 0.25 has to be avoided, if the full
equation of motion including precession is used. 

\section{Pulse-noise approach for many-spin systems\label{sec:Many-spin-systems}}

Many-spin systems usually have their own non-trivial dynamics, only
slightly modified by the coupling to the bath. Dynamic quantities
such as relaxation rates are typically due to spin-spin interactions.
The role of the coupling to the bath is merely to maintain the spin
system at the preset temperature. Thus the coupling to the bath $\lambda$
can be chosen small, so that the noiseless evolution time $\Delta t$
in the pulse-noise model can be made long, while satisfying $\varLambda_{N}\Delta t=2\gamma\lambda T\Delta t\ll1$.
This reduces the fraction of the computer time used to generate random
numbers to insignificant values, and the computation acquires the
speed of those for isolated systems. Of course, one has to generate
many random numbers for a good statistical averaging. In large systems
it occurs automatically because of a large number of spins. 

There are, however, special sutuations where coupling to the bath
becomes more important (\textit{non-precessional case}). This happens
for simple spin systems having integrals of motion that are broken
by the coupling to the bath (e.g., isotropic and uniaxial spin systems).
In particular, the prefactor in the overbarrier thermal-activation
rate of a magnetic particle with a uniaxial anisotropy is proportional
to $\lambda$, while adding a transverse field or a transverse anisotropy
breaks conservation of $S_{z}$ and makes the prefactor independent
of $\lambda$ and much larger \cite{garkencrocof99pre}. 

As time dependence of the integrals of motion is entirely due to spin-bath
relaxation and noise, one can discard the fast motion (precession
around the effective field) in Eq.$\,$(\ref{eq:LLL}). Resulting
\textit{slow} equation of motion can be solved with a much larger
integration step $\delta t$, saving computer time ($\gamma\lambda H_{\mathrm{eff}}\delta t\ll1$
instead of $\gamma H_{\mathrm{eff}}\delta t\ll1$, that makes a big
difference for weak damping $\lambda\ll1$). 

\begin{figure}
\begin{centering}
\includegraphics[width=8cm]{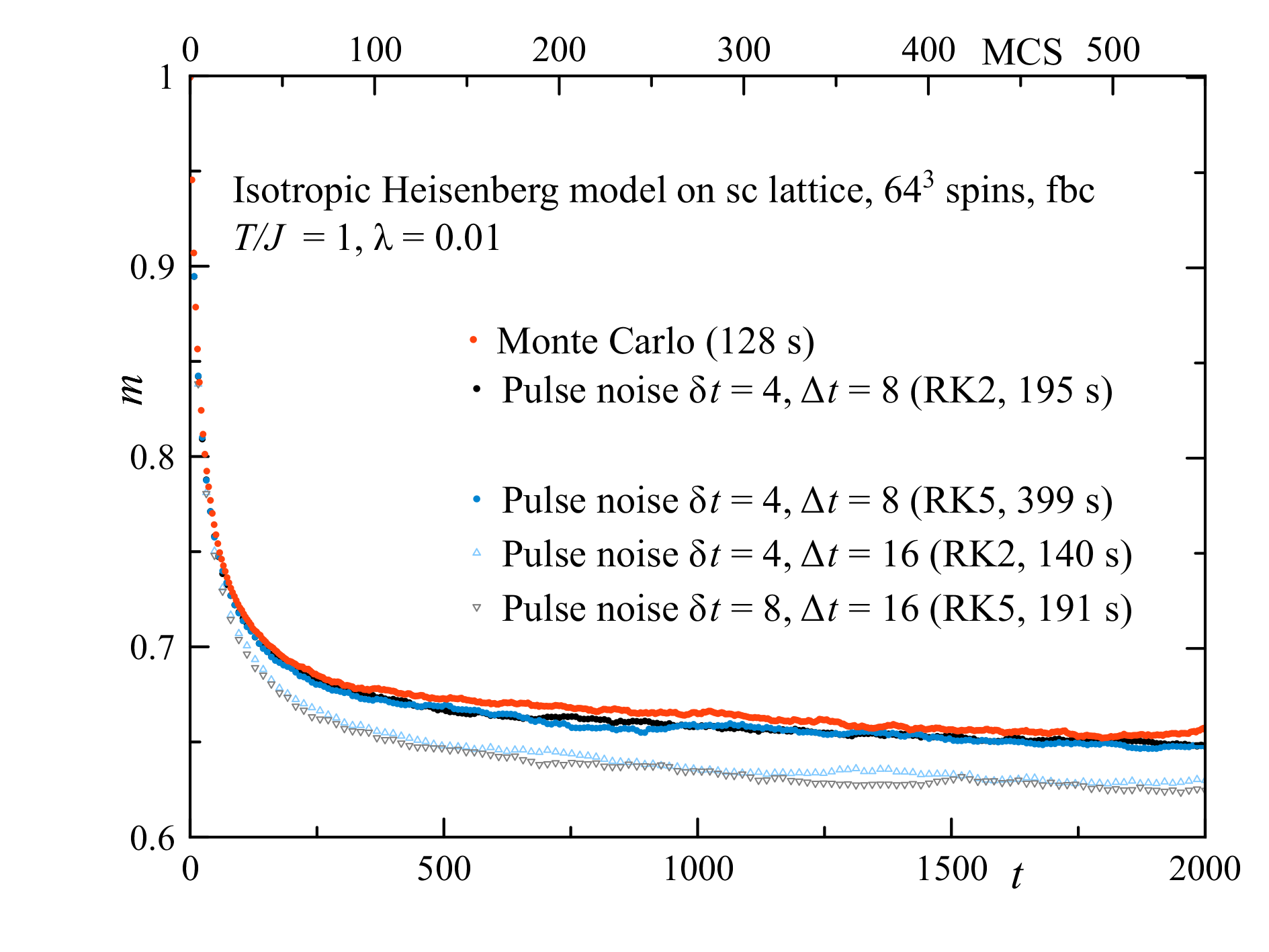}
\par\end{centering}
\caption{Monte Carlo vs pulse-noise approach: Relaxation of the magnetization
of a $64\times64\times64$ particle with isotropic Heisenberg coupling
and free boundary conditions out of a completely ordered state at
$T/J=1$. Numerical methods and computing times are indicated in brackets.}

\label{Fig_m_vs_t_Nx=00003DNy=00003DNz=00003D64_NSpins=00003D262144_T=00003D1}
\end{figure}
Fig.$\,$\ref{Fig_m_vs_t_Nx=00003DNy=00003DNz=00003D64_NSpins=00003D262144_T=00003D1}
shows the magnetization relaxation out of a fully ordered state for
a particle of $64^{3}=262144$ spins on a simple cubic lattice, coupled
by the isotropic Heisenberg exchange $J$ with free boundary conditions
at temperature $T/J=1$ that is below the Curie temperature $T/J=1.444$
in the bulk. The curves were obtained by variations of the pulse-noise
method and by the standard Metropolis Monte Carlo method \cite{metetal54jcp},
for a comparison. In the current implementation, one Monte Carlo step
(MCS) includes a successive update of all spins in the system by adding
a randomly directed vector $\mathbf{R}$ to each spin, and then normalizing
the spin, and computing the energy change $\Delta E$. This trial
is accepted unconditionally if $\Delta E<0$ and accepted with probability
$e^{-\Delta E/T}$ if $\Delta E>0$. The length of $\mathbf{R}$ was
chosen as $R=1.55\sqrt{T/(6J)}$ that yields about 50\% acceptance
rate in a wide temperature range. More details of the Monte Carlo
method and more trial choices can be found in Ref.$\,$\cite{now07Springer}.
Note that for such large system sizes, most of fluctuations self-average
and relaxation curves become pretty smooth without avegaring over
runs. The figure shows the results of one run for each set of parameters.
Both pulse-noise and Monte Carlo routines were not explicitly parallelized
in this numerical experiment.

In the pulse-noise method, precession terms have been discarded in
the equation of motion, that did not change the relaxation curve.
Dropping precession terms allowed a much larger integration time step
$\delta t$. However, the noiseless interval $\Delta t$ cannot exceed
8, as can be seen in the figure. For $\Delta t=16$ the applicability
condition of the method, Eq.$\,$(\ref{eq:phi_condition}), is violated
and the magnetization values fall visibly below those obtained by
the Monte Carlo. Correspondingy, $\delta t$ cannot be made large
enough to make high-order numerical integration methods win over low-order
methods. The results for $\delta t=4$ and $\Delta t=8$ obtained
by the RK2 midpoint routine are the same as those obtained by RK5
but the computation time is about two times shorter. Sensitivity of
the computation time to the ODE solver indicates that most time is
being spent on integration of noiseless equations of motion. 

Although these computations have been done for the particular damping
value $\lambda=0.01$, one can figure out the computation parameters
for any other value of $\lambda$, since in the precessionless case
$\lambda$ can be scaled out of the equations of motion. The efficiency
of the pulse-noise method in the precessionless case is the same for
any $\lambda$. 

To compare the real dynamics of the system with Monte-Carlo pseudo-dynamics,
one has to find a relation between time $t$ and the MCS \cite{nowchaken00prl,chubykaloetal03prb,now07Springer}.
Here it was done empirically by plotting the curves using dual axes
and adjusting the $t$ and MCS scales so that the relaxation curves
superimpose. Here, $t=2000$ corresponds to 550 Monte Carlo steps.
The speed of Monte Carlo is only slightly higher than that of the
pulse-noise method with $\delta t=4$ and $\Delta t=8$ using RK2.
To the contrary, Ref.$\,$\cite{evansetal14jpc} reports a 20 speed
advantage of the Monte Carlo in comparison to the standard stochastic
dynamics method using the Heun solver. 

It has to be added that the Monte Carlo routine can be parallelized
by splitting the particle into parts that can be processed in parallel.
This brings a significant speed gain, especially for large particles.
The ODE solvers used in the pulse-noise routines were written in the
vector form without explicit parallelization. In such cases Mathematica
is doing some parallelization at the processor level using Intel's
Math Kernel Library (MKL). Thus, the speed comparison above is somewhat
skewed to the favor of the pulse-noise method. The performance of
the Monte Carlo still can be improved by explicit parallelization.
However, this explicit parallelization becomes a useless burden if
statistical averaging over runs is performed. In this case one can
do many runs of the non-parallelized problem in parallel cycles, making
a better use of the multi-core processor. 

In any case, Monte Carlo is unbeatable in finding equilibrium states
of many-body systems at finite temperatures. The pulse-noise approach
in the precessionless case has a computation speed comparable with
that of Monte Carlo for equilibrium problems, as shown above. Its
advantage is in its universality \textendash{} the ability to deal
with real-time dynamics in addition to statics.

\section{Summary}

It was shown that replacing the continuous white noise acting on classical
spins by a pulse noise acting with a periodicity $\Delta t$ is superior
to the conventional method replacing the continuous noise by the rectangular
noise, constant within the intervals $\varDelta t$. The pulse-noise
approach leads to a considerable speed-up of numerical calculations
in the relevant underdamped case $\lambda\ll1$, since the maximal
possible value of $\Delta t$ that still ensures a good accuracy scales
with the relaxation time proportional to $1/\lambda$. Here one can
use high-order numerical integrators with a larger time step limited
by precession terms in the equation of motion. In this case $\delta t\ll\Delta t$
ensures a negligible contribution of noise-related operations into
computing time. 

In the cases where precession of spins can be discarded, time integration
step $\delta t$ can be increased up to $\Delta t$ that leads to
a further speed-up. Since here $\delta t$ is limited by $\Delta t$,
it cannot be made large enough to justify using high-order ODE solvers,
hence simpler second-order solvers work faster with a comparable accuracy.
Note that discarding precession terms is inefficient within the standard
stochastic formalism using the rectangular-noise approximation, since
still there is the noise-generated precession term that does not allow
a large increase of the time integration step. 

\section*{Acknowledgments}

This work has been supported by Grant No. DE-FG02- 93ER45487 funded
by the US Department of Energy, Office of Science.

\section*{Appendix: Details of numerical implementation}

All numerical calculation were done with Wolfram Mathematica using
compilation. For one-spin models, statistical averaging over realizations
of the noise (runs) were performed in parallel cycles on multi-core
computers. For the many-spin system in Sec. \ref{sec:Many-spin-systems},
single runs were performed, since the results self-average for large
systems. No explicit parallelization was done in this case. Mathematica
generates normal distribution with the \textit{Box-Muller} algorithm
from uniformly distributed real numbers. In parallel computations,
the latter are by default generated by \textit{Parallel Mercenne Twister}
due to Matsumoto and Nishimura. 

As the main ODE solver, Butcher's 5th-order Runge-Kutta (RK5) method
making six function evaluations per step was used. This method is
superior to the classical 4th-order Runge-Kutta method. Below is the
list of different numerical integrators for the equation $\dot{x}=f(t,x)$
with $\delta t\equiv h$ that were used in this project.

Heun (RK2) method
\begin{eqnarray}
K_{1} & = & hf[t,x]\nonumber \\
K_{2} & = & hf[t+h,x+K_{1}]\nonumber \\
x & = & x+\frac{1}{2}(K_{1}+K_{2}).
\end{eqnarray}

RK2 midpoint method
\begin{eqnarray}
K_{1} & = & hf[t,x]\nonumber \\
K_{2} & = & hf[t+\frac{1}{2}h,x+\frac{1}{2}K_{1}]\nonumber \\
x & = & x+K_{2}.
\end{eqnarray}

Butcher's RK5 method
\begin{eqnarray}
K_{1} & = & hf[t,x]\nonumber \\
K_{2} & = & hf[t+\frac{1}{4}h,x+\frac{1}{4}K_{1}]\nonumber \\
K_{3} & = & hf[t+\frac{1}{4}h,x+\frac{1}{8}(K_{1}+K_{2})]\nonumber \\
K_{4} & = & hf[t+\frac{1}{2}h,x-\frac{1}{2}K_{2}+K_{3}]\nonumber \\
K_{5} & = & hf[t+\frac{3}{4}h,x+\frac{3}{16}(K_{1}+3K_{4})]\nonumber \\
K_{6} & = & hf[t+h,x+\frac{1}{7}(-3K_{1}+2K_{2}+12(K_{3}-K_{4})+8K_{5})]\nonumber \\
x & = & x+\frac{1}{90}(7K_{1}+32K_{3}+12K_{4}+32K_{5}+7K_{6}).
\end{eqnarray}
For one-spin systems, the quantities in the formulas above are arrays
with one index, the spin component. For the many-spin system in Sec.
\ref{sec:Many-spin-systems}, they are arrays with four indices: the
spin component index and three lattice indices. Because of the vectorization,
the program implementations for one-spin and many-spin systems look
very similar.

\bibliographystyle{apsrev4-1}
\bibliography{C:/BIBLIOTHEK/gar-general,C:/BIBLIOTHEK/gar-own,C:/BIBLIOTHEK/gar-books,C:/BIBLIOTHEK/gar-relaxation,C:/BIBLIOTHEK/gar-spin,C:/BIBLIOTHEK/gar-tunneling,C:/BIBLIOTHEK/gar-oldworks,C:/BIBLIOTHEK/gar-superparamagnetic}
 
\end{document}